\documentclass[11pt]{article}
\pdfoutput=1
\usepackage{amsmath, amsthm, amssymb,xcolor,fullpage,url,booktabs}  %cite
\usepackage{graphicx}
\usepackage{algorithm}
\usepackage[noend]{algpseudocode}
\usepackage{subcaption}
\usepackage{authblk}
\ifdefined\blackandwhite
\usepackage[pdftex]{hyperref} %pagebackref
\else
\usepackage[colorlinks,pdftex]{hyperref} %pagebackref
\hypersetup{citecolor = blue, linkcolor = red!70!black}
\fi

\makeatletter

\newcommand{\algorithmicinput}{\textbf{Input:}}
\newcommand{\algorithmicresources}{\textbf{Resources:}}
\newcommand{\algorithmicgoal}{\textbf{Goal:}}
\newcommand{\algorithmicoutput}{\textbf{Output:}}
\newcommand{\Input}{\item[\algorithmicinput]}
\newcommand{\Resources}{\item[\algorithmicresources]}
\newcommand{\Goal}{\item[\algorithmicgoal]}
\newcommand{\Output}{\item[\algorithmicoutput]}
\algnewcommand{\LineComment}[1]{\Statex \hskip\ALG@thistlm \qquad\ \quad #1}
\makeatother

\newcommand{\nc}{\newcommand}
\nc{\rnc}{\renewcommand}

\newcommand{\bra}[1]{\left\langle #1\right|}
\newcommand{\ket}[1]{\left|#1\right\rangle}

\newcommand{\vev}[1]{\left\langle #1\right\rangle}

\def\be#1\ee{\begin{equation}#1\end{equation}}
\def\bea#1\eea{\begin{eqnarray}#1\end{eqnarray}}
\def\beas#1\eeas{\begin{eqnarray*}#1\end{eqnarray*}}
\def\ba#1\ea{\begin{align}#1\end{align}}
\def\bas#1\eas{\begin{align*}#1\end{align*}}
\def\bpm#1\epm{\begin{pmatrix}#1\end{pmatrix}}
\nc{\non}{\nonumber}
\nc{\nn}{\nonumber}
\nc{\eq}[1]{(\ref{eq:#1})}
\nc{\eqs}[2]{(\ref{eq:#1}) and (\ref{eq:#2})}
\rnc{\L}{\left} 
\nc{\R}{\right}
\nc{\ra}{\rightarrow}
\nc{\ot}{\otimes}
\nc{\grad}{{\vec{\nabla}}}

\newtheorem{thm}{Theorem}
\newtheorem*{thm*}{Theorem}

\newtheorem{proto}{Protocol}

\theoremstyle{definition}

\newtheorem{dfn}[thm]{Definition}
\theoremstyle{plain}

\makeatletter
\newtheorem*{rep@theorem}{\rep@title}
\newcommand{\newreptheorem}[2]{%
\newenvironment{rep#1}[1]{%
 \def\rep@title{#2 \ref{##1} (restatement)}%
 \begin{rep@theorem}}%
 {\end{rep@theorem}}}
\makeatother

\newreptheorem{thm}{Theorem}
\newreptheorem{lem}{Lemma}

\nc\eps{\epsilon}

\nc\cA{\mathcal{A}}
\nc\cB{\mathcal{B}}
\nc\cC{\mathcal{C}}
\nc\cD{\mathcal{D}}
\nc\cE{\mathcal{E}}
\nc\cF{\mathcal{F}}
\nc\cG{\mathcal{G}}
\nc\cH{\mathcal{H}}
\nc\cI{\mathcal{I}}
\nc\cJ{\mathcal{J}}
\nc\cK{\mathcal{K}}
\nc\cL{\mathcal{L}}
\nc\cM{\mathcal{M}}
\nc\cN{\mathcal{N}}
\nc\cO{\mathcal{O}}
\nc\cP{\mathcal{P}}
\nc\cQ{\mathcal{Q}}
\nc\cR{\mathcal{R}}
\nc\cS{\mathcal{S}}
\nc\cT{\mathcal{T}}
\nc\cU{\mathcal{U}}
\nc\cV{\mathcal{V}}
\nc\cW{\mathcal{W}}
\nc\cX{\mathcal{X}}
\nc\cY{\mathcal{Y}}
\nc\cZ{\mathcal{Z}}

\nc\bbC{\mathbb{C}}

\nc\bbF{\mathbb{F}}
\nc\bbM{\mathbb{M}}
\nc\bbN{\mathbb{N}}
\nc\bbR{\mathbb{R}}
\nc\bbZ{\mathbb{Z}}

\nc\benum{\begin{enumerate}}
\nc\eenum{\end{enumerate}}
\nc\bit{\begin{itemize}}
\nc\eit{\end{itemize}}

\nc{\todo}[1]{\textcolor{red}{todo: #1}}
\nc{\Anote}[1]{\textcolor{red}{Aram note: #1}}

\def\begsub#1#2\endsub{\begin{subequations}\label{eq:#1}\begin{align}#2\end{align}\end{subequations}}
\nc\qand{\qquad\text{and}\qquad}
\nc\mnb[1]{\medskip\noindent{\bf #1}}

\nc{\pder}[2]{\frac{\partial {#1}}{\partial {#2}}}
\nc{\p}{\partial}
\nc\bb{\boldsymbol{\beta}}
\nc\bg{\boldsymbol{\gamma}}
\nc\bt{\boldsymbol{\theta}}
\nc\smhalf{\mbox{$\frac 12$}}
\nc\smfrac[2]{\mbox{$\frac{#1}{#2}$}}

\begin{document}

\title{Quantum Algorithms for Fixed Qubit Architectures}
\author{Edward Farhi,$^{1, 2}$ Jeffrey Goldstone,$^{2}$ Sam Gutmann, Hartmut Neven{$^{1}$}}
\affil{$^{1}${Google Inc.\\ Venice, CA 90291}\\
\vspace{1ex}
$^{2}${Center for Theoretical Physics\\ Massachusetts Institute of Technology\\ Cambridge, MA 02139}}
%\date{$^{*}${Google Inc.\\ Venice, CA 90291}\\[1ex]
%$^{**}${Center for Theoretical Physics\\ Massachusetts Institute of Technology, Cambridge, MA 02139}}

%%%\googleaffiliation and \mitaffiliation
%%\author{Jeffrey Goldstone}
%%%\mitaffiliation
%%\author{Sam Gutmann}
%%\author{Harmut Neven}
%%\googleaffiliation
\date{}

\maketitle

\begin{abstract}
Gate model quantum computers with too many qubits to be simulated by available classical computers are about to arrive. We present a strategy for programming these devices without error correction or compilation. This means that the number of logical qubits is the same as the number of qubits on the device. The hardware determines which pairs of qubits can be addressed by unitary operators.  The goal is to build quantum states that solve computational problems such as maximizing a combinatorial objective function or minimizing a Hamiltonian. These problems may not fit naturally on the physical layout of the qubits. Our algorithms use a sequence of parameterized unitaries that sit on the qubit layout to produce quantum states depending on those parameters.  Measurements of the objective function (or Hamiltonian) guide the choice of new parameters with the goal of moving the objective function up (or lowering the energy). As an example we consider finding approximate solutions to MaxCut on 3-regular graphs whereas the hardware is physical qubits laid out on a rectangular grid. We prove that the lowest depth version of the Quantum Approximate Optimization Algorithm will achieve an approximation ratio of at least 0.5293 on all large enough instances which beats random guessing (0.5). We open up the algorithm to have different parameters for each single qubit $X$ rotation and for each $ZZ$ interaction associated with the nearest neighbor interactions on the grid. Small numerical experiments indicate that an enveloping classical algorithm can be used to find the parameters which sit on the grid to optimize an objective function with a different connectivity. We discuss strategies for finding good parameters but offer no evidence yet that the proposed approach can beat the best classical algorithms. Ultimately the strength of this approach will be determined by running on actual hardware.
\end{abstract}

%\cite{1,2,3,4,5,6,7,8}.

\section{Introduction}

We are entering an era when various experimental groups are building gate model quantum computers. In the near future these devices will have tens of qubits and soon many more \cite{1,2,3,4,5,6,7,8} A characteristic of hardware is that the qubits are positioned in space and the allowed direct two qubit gates are constrained by the architecture. \cite{1, 9, 10} However quantum algorithms are often written as a sequence of one and two qubit gates where the two qubit gates are between any pair of qubits. The algorithm designer can work in an idealized setting where all qubits are coupled to all others. To be run on hardware, the idealized quantum code must be compiled so that a two qubit gate between qubits which are physically separated is written as a sequence of gates involving a path through the hardware that comprises only two qubit gates that are physically coupled. This means that the depth of the compiled circuit can be much greater than the depth of the idealized circuit. With limited coherence times this may be a drawback.

Recently there has been interest in using quantum computers for approximate combinatorial optimization \cite{11,12} and for simulating quantum systems \cite{13,14,15,16}. In the first case the quantum computer is used to produce a quantum state that is dominated by computational basis states with a high value of some classical objective function. We can think of the objective function $C$ as a  sum over individual terms:

\be C(z) = \sum_\alpha C_\alpha(z)\ee
where each $C_\alpha$ acts on a small subset of the bits and has the value 1 for certain assignments of those bits and the value 0 on the other assignments. In the second case the quantum computer is used to produce a quantum state whose energy is near the ground state energy of a given Hamiltonian $H$. In both cases the ingredients are an initial state, $\ket{\text{initial}}$, and a sequence of unitary transformations that act on the initial state to produce a quantum state that depends on the parameters that define the sequence of unitaries. With $L$ unitaries we can write
\be |\boldsymbol{\theta}\rangle = 
U_L (\theta_L) \cdots  U_1(\theta_1) \ket{{\text{initial}}} \label{2}
\ee
where $\boldsymbol{\theta}$ denotes the collection $\theta_1,\ldots,\theta_L$ and  each $U_a$ depends on set of parameters $\theta_a$.
The goal is then to choose the parameters, $\boldsymbol{\theta}$, so that the expected value of the objective function $C$ 

\be f_L (\bt) =  \bra{\bt}C\ket{\bt}\ee
is big or that the expected value of the Hamiltonian $H$ 

\be E_L(\bt )= \bra{\bt}H\ket{\bt}
\label{eq:4}\ee
is small.

For the present discussion we will focus on combinatorial optimization and will briefly come back to  the quantum simulation case later. In all algorithms that we know of, the unitaries that drive the evolution depend on the objective function. For example the Grover algorithm \cite{17} is an alternation of objective function calls and objective-function-independent transformations.  In the case of the Quantum Approximate Optimization Algorithm \cite{11} the sequence of unitaries is an alternation of operators that depend on the objective function $C$ and those that do not. In the case of the Quantum Adiabatic Algorithm \cite{18} there is continuous time evolution governed by a Hamiltonian that depends on the objective function $C$ and (\ref{2}) can be viewed as a ``Trotterized'' approximation to this evolution where roughly half of the unitary operators depend on the objective function $C$. In these situations the objective function $C$ is playing a dual role. It is the object to be optimized and also an ingredient in the unitaries that govern the evolution.

In this paper we begin to explore the possibility of decoupling the objective function from the unitaries that are used to produce the quantum state.  It suffices for this discussion to consider only objective functions that can be written as a sum of individual terms where each $C_\alpha$ involves only two bits. But the connectivity of the objective function may have nothing to do with the pairwise connectivity of the hardware. For the unitaries that appear in (\ref{2}) we propose using the toolkit of unitaries that are given to us by the hardware. The hardware architecture  determines which qubit pairs can be acted on by the two qubit unitaries. And also the form of the two qubit unitaries should be dictated by hardware limitations. However for most experimental implementations the range of possible two qubit unitaries is large and typically any one qubit unitary can be implemented. Note that when writing the objective function there is an implicit ordering of the qubits. The qubits on the hardware can also be labelled from 1 to $n$, the number of qubits. We have the freedom to decide how to match these two sets and we will attempt to make judicious choices.

Imagine that we have an $n$ qubit quantum computer and can make the states of the form (\ref{2}) where we have applied $L$ unitaries depending on the parameters $\bt$. Here we give pseudo code for how to run the quantum computer attempting to find a large value of $C$. The algorithm will generate a sequence of parameters $\bt$ and associated estimates of $f_L(\bt)$. We require an enveloping classical algorithm that given such a sequence will pick new values of $\bt$ that tend to move $f_L(\bt)$ to higher values. For now assume that such an algorithm is at hand. It might involve gradient ascent or some gradient independent method (see for example \cite{2017arXiv170101450G}). 

\begin{algorithm}[t]
\caption*{{\bf Pseudo Code for a Generic Variational Quantum Algorithm\\ Applied to Combinatorial Optimization}}\label{qoa}
\begin{algorithmic}[1]
\Input
 A classical objective function $C$ on $n$ bit strings that can efficiently be evaluated on any input string $z$ 
%2) A classical optimization routine that takes as input a sequence of parameters ${\bf \Theta}$ and associated values $f_L(\bt)$ and outputs a new $\bt$.
\Resources  
  \Statex 1.  An $n$ qubit quantum computer that can produce states of the form (2)
   \Statex     2.  A classical optimization routine that takes as input a sequence of parameters $\bt$ and associated values $f_L(\bt)$ and outputs a new value of $\bt$
%{\bf Goal:} A string $z$ that has a ``large'' value of $C(z)$
\Goal A string $z$ that has a ``large'' value of $C(z)$
\Require
\Statex Pick a repetition number $R$
\Statex Pick a stopping criterion which may depend on the quality of the objective function value or on the number of times the quantum computer has been called
\Statex Pick an initial state $\ket{\text{initial}}$
\Statex Pick one of the $n!$ ways of assigning the $n$ bits associated with the objective function to the $n$ qubits on the hardware
\Statex Pick an initial set of parameters $\bt$
\Statex {\bf while} the stopping criterion is not satisfied
\Statex {\quad\ {\bf for} $R$ times} 
		\LineComment Run the Quantum Computer and make the state $\ket{\bt}$
		\LineComment Measure in the computational basis: this produces a string $z$
		\LineComment Evaluate $C(z)$
%\EndFor
\Statex { \quad\ \bf end for}
\Statex \quad\ Average the $R$ values of $C$ to get an estimate $\widehat{f}_L(\bt)$ of $f_L(\bt)$
\Statex \quad\ Use the previous values of $\bt$ and $\widehat{f}_L(\bt)$ to choose a new value of $\bt$
%\EndWhile
\Statex {\bf end while}
\Output  The string $z$ seen during the {\bf Procedure} with the highest value of $C(z)$ %\Statex{\bf Output:} 
%\EndProcedure
\end{algorithmic}
\end{algorithm}

%Input: A classical objective function $C$ on n bit strings that can efficiently be evaluated on any input string z
%
%Goal: Find a string $z$ that has a ``large'' value of $C(z)$
%
%Pick a repetition number $R$
%
%Pick a stopping criterion which may depend on the quality of the objective function value or on the number of times the quantum computer has been called
%
%Pick an initial state $\ket{\text{initial}}$
%
%Pick one of the $n!$ ways of assigning the $n$ bits associated with the objective function to the $n$ qubits on the hardware
%
%Pick an initial set of parameters $\bt$
%
%While the stopping criterion is not satisfied
%
%			Repeat $R$ times
%
%			Run the Quantum Computer and make the state $\ket{\bt}$
%
%			Measure in the computation basis. This produces a string $z$.
%
%			Evaluate $C(z)$
%
%			End Repeat
%
%	Average the $R$ values of $C$ to get an estimate of $f_L(\bt)$
%
%	Use the previous values of $\bt$ and $f_L(\bt)$ to choose a new value of $\bt$.
%
%End While

\begin{figure}[H]
  \centering
  {\includegraphics[width=\textwidth]{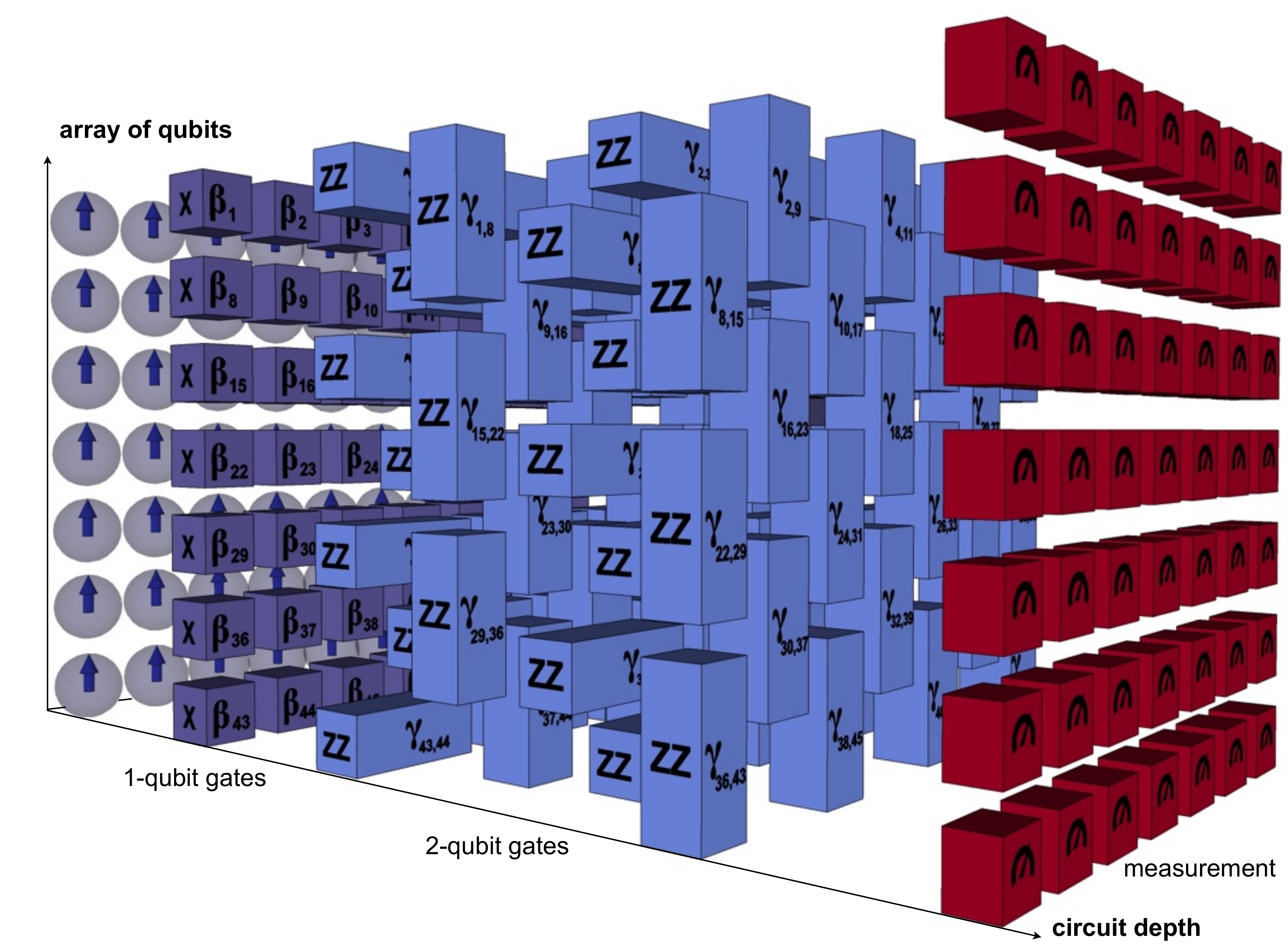}}
  \caption{Space-time volume of a quantum circuit computation. An array of qubits, here arranged on a 7 by 7 grid, is initialized by preparing the quantum state of each qubit. Then a layer of 1-qubit gates and a layer of 2-qubit gates act on the qubits. In our case we apply 1-qubit gates of the form $ \exp (- i  \beta_j  X_j )$ followed by a round of 2-qubit gates of the form $\exp ( i  \gamma_{jk}  Z_j Z_k  )$. Note that due to hardware restrictions it may be necessary to apply the 2-qubit gates in several rounds as depicted, as it may not be possible for one qubit to participate in two gates simultaneously. For simplicity the figure only shows one round of 1- and 2-qubit gates but many rounds can be applied. However in pre-error corrected processors the circuit depth is limited by the imperfect fidelity of the gates. The final step in the circuit consists of measuring the qubits.}
\label{Figure1}
\end{figure}

It is tempting to view a variational quantum circuit on a fixed connectivity graph as a ``quantum neural network." In our quantum case, as in classical machine learning, we are interested in generating output states with desirable properties and in both cases the output state depends on a set of parameters that are chosen during training via classical optimization. The connectivity graph remains fixed. In both the quantum and classical case we want to be assured that good output states are attainable for suitably chosen parameters but we expect that the more free parameters there are the more iterations it will take to optimize them.  However since the goal is approximate optimization the computational effort to achieve a good solution with more parameters may be less than is required with fewer parameters. For producing the entangled ground state of a Hamiltonian $H$, it is natural to expect that a quantum computer will be more efficient. This is the situation in quantum simulation and one can imagine the application of variational quantum circuits to chemistry, materials research, field theory or nuclear physics. In the case that we work with classical data, for instance when we want to solve an optimization problem or prepare a probability distribution it is less obvious  where the advantage of a quantum neural network would come. Our hope is that the presence of entangling gates enables interference along the computational path which allows the preparation of desirable output states with shallower circuits or less training effort than would be possible with just classical gates. As work on quantum supremacy has shown we do know that quantum circuits can prepare probability distributions that are out of reach for classical machines \cite{19,20,21,22,23,24,25,26,27,28,29,30}. Even though many approaches to quantum machine learning have been proposed \cite{31} quantum neural networks as discussed here are in a way minimal in that they directly optimize the native gate library of a quantum processor.   

%We can view machine learning as process that produces a probability distribution over n bit strings. The goal is to optimize some characteristic of this probability distribution. The computer has a set of parameters that control the distribution. For example in a Boltzman Machine the distribution is a thermal distribution over classical spins (bits) using an Ising model for the energy function. The parameters control the coupling strengths. The architecture of the connections may beindependent of the properties of the desired distribution. Still with enough parameters and training, the parameters can be set to give desired distributions. In thissense our proposal for how to run a quantum computer is analogous. We build a quantum state that depends on parameters through the unitaries that are given to us by the hardware team. Our goal is to set the parameters such that the final quantum state has certain desirable properties. Our hope is that thedistributions available in the quantum setting have a richness that may result in our being able to achieve desired properties with less computational cost than what isrequired in the classical setting.

\section{QAOA on a Native Hardware Graph}

At first glance it may seem that using unitaries that do not depend on the objective function can not possibly produce quantum states with high values of the objective function. To address this we use as our example a specific qubit layout that is inspired by ongoing experimental efforts. We imagine that the $n$ qubits are arranged on a square grid with $\sqrt n$ qubits in each row and column. Except at the borders, each qubit is coupled to 4 others. A seven by seven example is shown in figure 1. We will stick with the grid throughout this paper but other layouts are of course possible.

The Quantum Approximate Optimization Algorithm was first applied to the combinatorial problem MaxCut on 3-regular graphs. Here the objective function is a sum of terms, one for each edge in the graph, with the value
\be C_{\langle ij\rangle} (z) = \mbox{$\frac 12$} ( 1- z_i z_j) \label{eq:edge-term}\ee
for the edge connecting $i$ to $j$ and $z_i = \pm 1$ and accordingly 
\be C(z) = \sum\limits_{\langle ij\rangle \text{in the graph}} C_{\langle ij\rangle} (z). \label{eq:6}  \ee
We then have the unitary operator, diagonal in the computational basis,
\be U( C, \gamma)\ket{z} = \exp (- i \gamma C(z)) \ket{z}  \ee
which depends on the single parameter $\gamma$. Here we are discussing the original version of the QAOA so the objective function appearing in $U$ corresponds to the 3-regular graph under consideration. There is also the operator
\be U ( B, \beta) = \exp ( - i \beta B) \ee
where
\be B = \sum_i X _i \qand  X_i = \sigma_x^i \ ,\ee
and the sum is over all qubits. The quantum computer is run to produce the state
\be \ket{\gamma, \beta, C} = U(B,\beta) U(C,\gamma) \ket{s}\label{eq:9}\ee
where
\be \ket{s} = \frac{1}{\sqrt{2^n}} \sum_z \ket{z}.\label{eq:10}\ee
We include the $C$ label in the ket to remind us that the state has an explicit dependence on the objective function $C$. 

The goal is to make         
\be 
\bra{\gamma, \beta, C}C
\ket{\gamma, \beta, C} \label{eq:11} \ee
large.  It was shown that there is a choice of $\gamma$ and $\beta$ such that the approximation ratio for this problem is at least 0.6924 for all instances of 3-regular MaxCut.  The approximation ratio is the value of \eqref{eq:11} divided by the maximum of $C$ over all strings.   The significance of this result is that it holds for all instances and it beats random guessing which gives an approximation ratio of 0.5. (Random guessing may seem like a low bar but until the introduction of the Goemans-Williamson algorithm \cite{32}, random guessing was the only algorithm with a provable worst case performance guarantee.) There are classical algorithms that give better performance guarantees than this quantum algorithm.  However this result is for the shallowest version of the quantum algorithm and performance improves with circuit depth so perhaps with sufficient depth the quantum algorithm will outperform the best classical.  

In the discussion of the previous paragraph the objective function $C$ appears in (\ref{eq:11}) both as the operator being evaluated and through its use in building the state.   Now we assume that the hardware gives us a graph, to be specific a square grid, and we want the state to be constructed using the same clauses living on the grid.  Let 
\be G (z) = \sum_{\langle i j\rangle \text{in the grid}}  \mbox{$\frac 12$} ( 1 - z_i  z_j ) \ee
and let the produced state be
\be \ket{\gamma,\beta,G}  = U(B, \beta)  U(G, \gamma)\ket s. \label{eq:13}\ee
We now want to make 
\be \bra{\gamma,\beta,G} C \ket{\gamma,\beta,G} \label{eq:14}\ee
as big as possible.  What we will now show is that we can achieve an approximation ratio of  0.5293 on all large enough instances of 3-regular MaxCut using the lowest depth version of the algorithm.  The significance of this result is that it establishes that we can achieve a non-trivial worst case approximation ratio even when the state is produced without direct use of the objective  function $C$.  We now show how this result is obtained.

Any 3-regular graph with $n$ vertices has $3n/2$ edges. The border of the grid has order $\sqrt{n}$ vertices and edges. We take $n$ large enough so that we can neglect the border of the grid and assume that each vertex has valence 4 and say that there are $2n$ edges.  Now the vertices of the grid are labeled from 1 to $n$ as are the vertices of the 3-regular graph.  Consider edge $\vev{ij}$ in the 3-regular graph.  We wish to evaluate 
\be (-\mbox{$\frac 12$})\bra{\gamma,\beta,G} Z_i Z_j \ket{\gamma,\beta,G}
\label{eq:contribution}\ee
which makes a contribution  to the expected value of the objective function $C$ in (\ref{eq:14}).   There are 4 cases to consider.
\begin{figure}[h!]
 	\begin{subfigure}[t]{0.5\textwidth}
	\centering
	\includegraphics{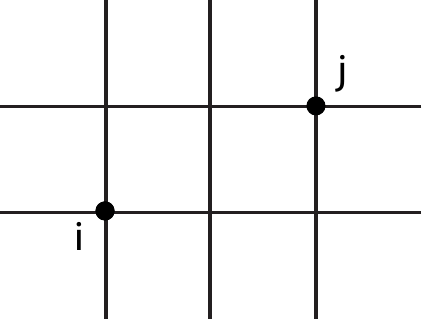}%[width=0.56\columnwidth]
	\caption*{\centerline{Figure 2.0: An example where}\\   \centerline{qubits i and j are more than 2 edges apart.}}
	\label{figure2.0}
	\end{subfigure}
	\begin{subfigure}[t]{0.5\textwidth}
	\includegraphics{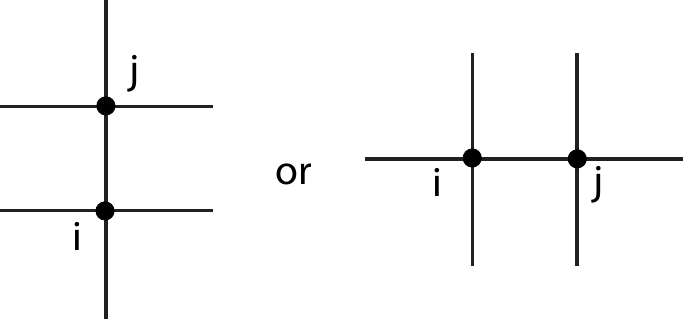}%[width=0.9\columnwidth]
	\caption*{Figure 2.1: Qubits i and j are one edge apart}		
	\label{figure2.1}
	\end{subfigure}
	\par\bigskip
	\par\bigskip
	\begin{subfigure}[t]{0.5\textwidth}
	\centering
	\includegraphics{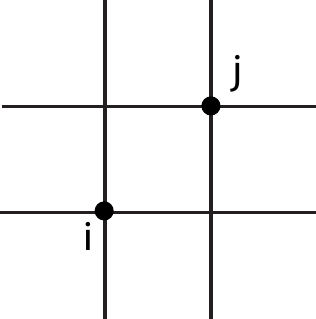}%[width=0.45\columnwidth]
	\caption*{\centerline{Figure 2.2:  An example  where qubits i and j}\\  \centerline{are one vertical and one horizontal edge apart}}	
	\label{figure2.2}
	\end{subfigure}\hspace{-2em}
	\begin{subfigure}[t]{0.5\textwidth}
	\includegraphics{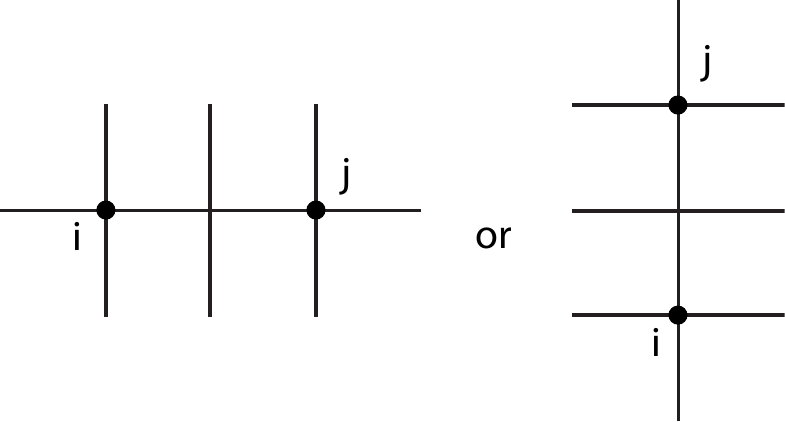}%[width=1.1\columnwidth]
	\caption*{\centerline{Figure 2.3: Qubits i and j are}\\ \centerline{two horizontal or two vertical edges apart}}		
	\label{figure2.3}
	\end{subfigure}
	\caption*{}
	\label{figure2}
\end{figure}

\begin{enumerate}
\addtocounter{enumi}{-1}
\item    $i$ and $j$ are separated on the grid by more than two edges. As an example see Figure 2.0.  In this case the expected value of \eq{contribution} is 0 which corresponds to a clause value of 0.5 as can be seen from (\ref{eq:edge-term}).  The calculations for this case and the next three are in appendix A.

\item $i$ and $j$ are separated on the grid by one vertical edge or by one horizontal edges. See Figure 2.1. In this case the expected value in (\ref{eq:contribution}) is
\be 
F_1 = \smhalf \sin 4\beta \sin \gamma \cos^3\gamma .
\label{fsmall}
\ee

\item $i$ and $j$ are separated on the grid by one vertical edge and one horizontal edge. See Figure 2.2.  Now the expected value in (\ref{eq:11}) is
\be F_2 = -\sin^2 2\beta \sin^2\gamma \cos^6 \gamma.\label{eq:17}\ee

\item  $i$ and $j$ are separated on the grid by two vertical edges or two horizontal edge. See Figure 2.3.   The expected value in (\ref{eq:contribution}) is
\be F_3 = -\smhalf \sin^2 2\beta \sin^2\gamma \cos^6 \gamma.\label{eq:18}\ee
\end{enumerate}

Suppose the number of edges from the 3-regular graph which end up in case 1 is $m_1$ and similarly  define $m_2$ and $m_3$.  Summing (\ref{eq:contribution}) over all edges gives
\be m_1 F_1 + m_2 F_2 + m_3 F_3.\ee
We want to choose $\gamma$ and $\beta$ to make this as positive as possible.   Just using calculus we see that the maximum occurs at
\be \beta= \frac{1}{4} \tan^{-1}\, \left(\frac{m_1}{m' \sqrt{\lambda}}\right) \qand \gamma = \pi/6\ee
where $\lambda = 27/256$ and $m' = m_2 + \smhalf m_3$, and the maximum equals
\be \frac 12  \left[ \sqrt{m_1^2 \lambda  + (m')^2\lambda^2} - m'\lambda\right].
\ee
 Note that for any $m_1>0$ this is greater than 0.  Now the expected value of the objective function $C$ in the state $\ket{\gamma,\beta,G}$ is
\be \frac m 2 +\frac 12  \left[ \sqrt{m_1^2 \lambda  + (m')^2\lambda^2} - m'\lambda\right],
\label{eq:22}
\ee
where $m$ is number of edges in the 3-regular graph, $3n/2$.   For the approximation ratio the denominator is upper bounded by the number of clauses, $m$, so the approximation ratio is at least
\be 
\frac 12 
+\frac{1}{2m}  \left[ \sqrt{m_1^2 \lambda  + (m')^2\lambda^2} - m'\lambda\right] . \label{eq:23}
\ee
To bound the approximation ratio away from 1/2 as $n$ gets big we need $m_1$ to scale with $m$.  Assume that the ordering of the vertices on the grid is fixed.  We now need to label the vertices on the 3-regular graph so that the number of edges that end up being type 1 scales with $n$.  We can achieve this with a simple greedy approach.

The algorithm we now give will produce a way of assigning the vertices of the 3-regular graph to the grid such that the number of type 1 edges in the grid is at least $n/2$. Start by choosing distinct vertices two at a time, $v_1$, $v_2$ then $v_3$, $v_4$, such that there is an edge in the 3-regular graph between the vertices in each pair.  At some point, after $k$ pairs have been chosen no further connected pairs can be found among $\left\{v_{2k+1}, ... v_n\right\}$.  (Possibly $k=n/2$ if there is a perfect matching and the process finds it.) Thus the remaining $(n-2k)$ vertices each have 3 edges terminating in $\{ v_1, \ldots v_{2k}\}$. Now pick two vertices $v_i$ and $v_j$ with $i\leq2k$ and $j>2k$ such that there is an edge between $v_i$ and $v_j$.  After $v_i$ and $v_j$ are paired they are not considered for further pairing. Of the $3(n-2k)$ edges shown on the right in figure 3 at most 4 are no longer available once $v_i$ is paired with $v_j$.  (In addition to the edge between $v_i$ and $v_j$ there are two more edges starting at $v_j$ and there may be one additional edge between some $v_l$, where $l>2k$, and $v_i$.)  So this pairing can be carried out at least $3(n-2k)/4$ times.  Adding the $k$ pairs previously found we have
\be
                                        k + 3(n-2k)/4  =  3n/4 -k/2
\ee
pairs which since $k\leq n/2$ is at least $n/2$.  We have identified a subgraph of all $n$ vertices and  at least $n/2$ edges of the the original 3-regular graph. This subgraph consists of disconnected segments of 2, 3 or 4 vertices and isolated vertices. This subgraph can easily be superimposed (in many ways) on a $\sqrt{n}$ by  $\sqrt{n}$ grid. A simple way to embed the segments is to start in the upper left hand corner of the grid and start placing the segments horizontally. When you reach the right border of the grid, if your current segment does not fit on the horizontal row, make a bend so that it ends up on the horizontal row one below the top. Now continue moving to the left and complete the zigzag pattern.  All the segments can be fit on the grid in this way. This guarantees that we can pick $m_1$ to be at least $n/2 = m/3$. (We don't claim this is best possible.)

\begin{figure}[t]
  \centering
%  {\includegraphics[width=0.5\columnwidth]{Fixed_Layout_Fig2.pdf}}
 {\includegraphics{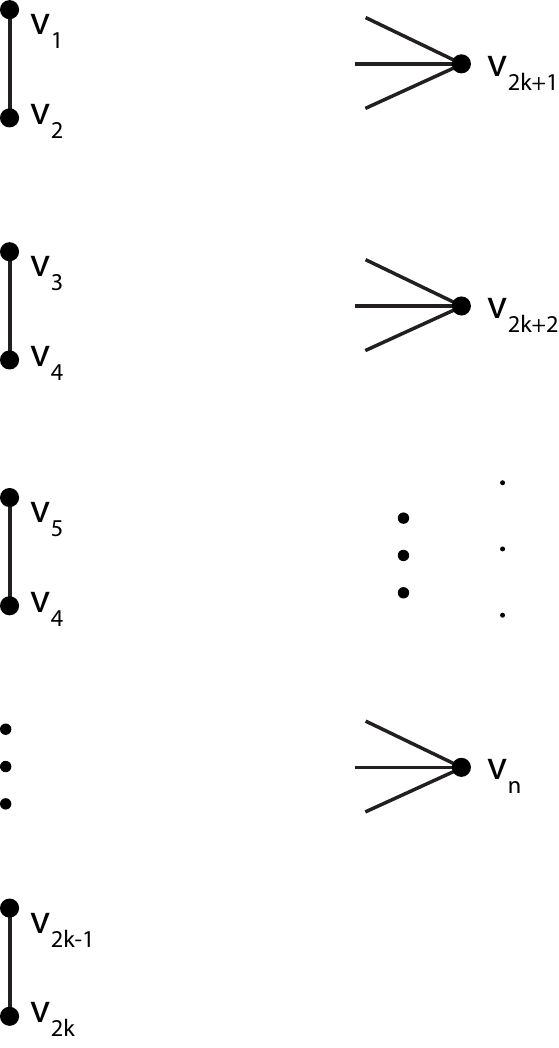}}%[scale=0.65]
  \caption{A representation of the greedy algorithm used to achieve a pairing of the bits}
\label{Figure3}
\end{figure}

%\begin{figure}[H]
%  \centering
%%  {\includegraphics[width=0.5\columnwidth]{Fixed_Layout_Fig2.pdf}}
% {\includegraphics{figure4.pdf}}%[scale=0.65]
%  \caption{Disconnected segments with 2, 3 or 4 edges and isolated vertices that can be placed on the grid. The greedy algorithm does not say how to place the segments on the grid and we illustrate choices such as horizontal, vertical or with bends. }
%\label{Figure3}
%\end{figure}

We now know that we can, with minimal computational cost, assign the vertices of the 3-regular graph to the vertices of the grid such that $m_1 \geq m/3$.  However the approximation ratio in \eqref{eq:23} decreases with $m'$.  This is because the contributions from type 2 and type 3 pairs (see \eqref{eq:17} and \eqref{eq:18}) come with negative signs.  So it might pay to arrange the assignment of the vertices on the grid to minimize the number of type 2 and type 3 pairs. Even without this extra effort we know that $m' \leq 2m/3$ since $m_1 + m_2 + m_3 \leq m$.  Picking $m_1=m/3$ and $m' = 2m/3$ we have that the approximation ratio in \eqref{eq:23} is at least 0.52938...  The error made at the border is of order $\sqrt{n}$ in \eqref{eq:22} and $1 / \sqrt{n}$ in \eqref{eq:23} so for $n$ large enough we have that the approximation ratio is at least 0.5293 for any 3-regular graph using the grid for the unitary that builds the quantum state as long as we assign the vertices as just described. This discussion is for the lowest depth version of the algorithm using only two angles $\beta$ and $\gamma$.

Here is a small numerical example. We generated a 20 bit 3-regular graph and ran the original QAOA at $p = 1$ to produce the state \eqref{eq:9}. Optimizing the two angles $\gamma$ and $\beta$ gave an approximation ratio of 0.7519. We then switched to a 4 by 5 grid and assigned the vertices of the graph to the grid using the procedure just given which produced 11 pairs. Optimizing the two angles in \eqref{eq:14} gave an approximation ratio of 0.6424. As expected the mismatch between the grid and the graph lessened performance.

What we showed sets a baseline for the idea that we can use a unitary that lives on the grid to solve a problem on an unrelated graph. Admittedly we lined the two up somewhat at a marginal computation cost.  This step was necessary because at the shallowest depth version of the algorithm, qubits that are separated by more than two edges on the grid do not sense each other. See case 0 above. However as the depth increases to $\sqrt{n}$, all qubits are entangled with each other and this type of preprocessing may not be crucial.

\section{Introducing Extra Parameters}

We now look at opening up the algorithm so that each unitary operator on individual qubits and each unitary on connected pairs on the grid has its own parameter.  We have the freedom to do this and it can only increase the best value of $C$ found. Also, since the edges in the grid have a complicated relationship to the edges in the parent graph it seems best to let the associated angles vary independently of each other and hope that the optimization routine "sees" the hidden structure. We could allow for arbitrary one and two qubit unitaries but to keep our discussion more manageable we will make choices for what is allowed.  For single qubits we will use
\be U_S ( \beta_j) = \exp (- i  \beta_j  X_j )\ee
so that each qubit has its own parameter $\beta_j$  which we can combine into 
$\bb = ( \beta_1, â\dots  \beta_n)$ and write 
\be U_S (\bb ) = \prod_j U_S (\beta_j) \ee
for the application of all the single qubit rotations at once.   For each pair of qubits $j$ and $k$ connected on the grid we introduce a parameter $\gamma_{jk}$ and then 
\be U_P ( \gamma_{jk}) = \exp ( i  \gamma_{jk}  Z_j Z_k  )\ee
which can be combined to make 
\be U_P ( \bg ) = \prod_{\langle jk\rangle \text{ in the grid}}
U_P ( \gamma_{jk})\ee
where $\bg$ has all of the individual $\gamma_{jk}$'s combined.  Now for a level-$p$ circuit we introduce a set of $p$ $\bb$'s, $\bb^{(1)},\ldots\bb^{(p)}$ and a set of $p$  $\bg$'s, $\bg^{(1)},\ldots\bg^{(p)}$ and the state
\be \ket{\bg^{(1)},\ldots \bg^{(p)}, \bb^{(1)}, \ldots \bb^{(p)}, G} = U_S (\bb^{(p)} ) U_P (\bg^{(p)} ) \cdots U_S ( \bb^{(1)}) U_P (\bg^{(1)}) \ket s . \label{eq:29}
\ee
The goal now is to find the parameters $\bg^{(1)},\ldots \bg^{(p)}, \bb^{(1)}, \ldots \bb^{(p)}$ for which the expected value of some objective function $C$ is large. Since $X_j$ and $Z_j Z_k$ have eigenvalues $\pm 1$, it suffices to let every $\gamma$ and every $\beta$ lie between $0$ and $\pi$.

\subsection{Making Computational Basis Cat States}\label{sec:3-1}
We now argue that with $p = \sqrt n$ we have enough freedom in (\ref{eq:29}) to make any computational basis state.  Actually since the starting state $\ket s$  in (\ref{eq:29}) is an eigenstate of 
$X_1  X_2  \ldots  X_n$ and this operator commutes with all of the unitaries we have just introduced, it follows that we can only produce states of the form
\be \ket w + \ket{\bar w} \label{eq:w-cat}\ee
where $\bar{w}$ is the bit flip of $w$.  This means that we are forced into producing states that are quantum entangled. We view this as a virtue since we want to stay away from purely classical computation.  

We start by making the usual cat state
\be \ket{00\ldots 00} + \ket{11\ldots 11}.\ee
First consider the two qubit operator
\be
U_{12} = e^{- i \frac{\pi}{4}} U_S \, \left(\beta_2 = - \tfrac{\pi}{4}\right) \, U_P \left(\gamma_{12} = \tfrac{\pi}{4}\right)\label{eqA1} %\tag{A1}
\ee
whose action is 
\be
U_{12} \ket{0+} = \ket{00} \label{eqA2} %\tag{A2}
\ee
\be
U_{12} \ket{1+} = \ket{1 1} \nonumber
\ee
where as usual $\ket{+} = \frac{\ket{0}+\ket{1}}{\sqrt{2}}$. This gives us
\be
U_{12} \ket{++} = \tfrac{1}{\sqrt{2}}\,  \left(\ket{00} + \ket{11} \right) . \label{eqA3}%\tag{A3}
\ee
Now append a third qubit in the $\ket{+}$ state to the cat state in \eqref{eqA3} and act with $U_{23}$ on the second and third qubits to get a 3 qubit cat state:
\be
U_{23} \tfrac{1}{\sqrt{2}}\, \left(\ket{00} + \ket{11}\right) \ket{+} = \tfrac{1}{\sqrt{2}}\, \left(\ket{000} + \ket{111}\right) \label{eqA4} %\tag{A4}
\ee
%First note that if we have two qubits each in a plus state $\ket + = \frac{\ket 0 + \ket 1}{\sqrt 2}$ that
%\be U_S(\beta_1=\smfrac{3\pi}{8}) U_S(\beta_2= \smfrac{3\pi}{8}) U_P(\gamma_{12} = \smfrac{\pi}{4} ) \ket{++} =  \frac{e^{-i3 \pi/4}}{\sqrt 2} \, \left(\ket{00} + \ket{11}\right).\ee
%
%Now append a third qubit in the $\ket +$ state and act with the operation just given but on the second and third qubits.  This will produce the state
%\be \frac{\ket{000} + \ket{111}}{\sqrt 2}.\ee
%%
Iterating this allows us to make a cat state on the grid with $p=\sqrt n$.   Start at the central qubit of the grid.  (Assume that $n$ is odd so there is a clear central qubit for simplicity.) Using a slight generalization of the above construction, at $p=1$, we can make a cat state of the central qubit and its 4 neighbors.  At the next level we can  entangle all qubits that are a distance 2 away from the central qubit.  The corners of the grid are $\sqrt n$ edges away from the center. So with this depth we can make a cat state.  Since $\exp ( - i  X  \pi /2 )$ is proportional to $X$ which flips bits, we can make any cat state of the form \eq{w-cat} with a slight modification of the procedure. 

Another way to try to make cat states of the form \eq{w-cat} for an arbitrary $w$ is to run the quantum optimization algorithm using a simple objective function that is maximized at $w$ and $\bar w$.  For example using the Hamming distance from $w$ we can take
\be C_{\text{quad}} ( z ) = - \text{Ham}(z,w)(n-\text{Ham}(z,w)) + (n/2)^2.\label{eq:35}\ee
Using the grid states (\ref{eq:29}) we search for angles that maximize $C_{\text{quad}}$.  The idea here is that the objective function is fairly simple and perhaps finding angles on the grid that maximize this objective function is not hard.  Again the unitary that produces the state is not associated with the objective function explicitly. 

\subsection{Numerical Example}
%We know that with depth $\sqrt{n}$ we can produce arbitrary computational basis states, more specifically we produce entangled states of the form (\ref{eq:w-cat})  This means that if we search for a large value of the objective function $C$, say it is $w$, we know that the state \eqref{eq:w-cat} is of the form \eqref{eq:29} for some set of parameters. This does not of course mean that the angles that bring us to this state are easy to find but at least we know they exist.  You can object here and say suppose we use for our one qubit operations the $Y$ operators and not the $X$ operators.  Acting on a $\ket +$ state with $\exp ( \pm i  Y \pi /4 )$ gives the computational basis states $\ket 0$ and $\ket 1$ depending on the sign. This means that if we just use a depth-one algorithm with rotations about $Y$, we can produce any computational basis state so the solution to our optimization problem is in the domain of this search space.  But this procedure uses only product states and is a essentially equivalent to replacing bits by classical rotating spins.  This approach finds optimal solutions for say NP hard problems and therefore presumably requires exponential time in worst case. Our hope, and it is really only a hope, is that the entangling  quantum procedure will allow us to find say good approximate solutions to  combinatorial search problems which evade our best classical algorithms.

We performed a small scale numerical experiment to demonstrate the strategy of opening up the parameter space as in \eqref{eq:29} to optimize an objective function. Working at 16 bits we tossed a 3-regular graph and used the MaxCut objective function given by \eqref{eq:6} although for the unitary we used \eqref{eq:29} which lives on the grid. The assignment of the vertices of the 3-regular graph to the vertices of the grid was random. (We did not bother with the preprocessing described in the previous section.) We chose $p$ to be $\sqrt{n}$ which is 4. The 4 by 4 grid has 24 edges so at each level there are 24 $\gamma$'s and 16 $\beta$'s for a total of 160 parameters.  To run the simulation we needed a classical optimizer to drive up the expected value of $C$ in the the state \eqref{eq:29}.  We used a home made Nelder Mead routine \cite{33} running Matlab on a laptop.  For the initial simplex the routine tosses 161 points in the $[0, \pi]^{160}$ cube.  The best of ten runs achieved an approximation ratio of 0.9399.   We contrast this with using an opened up set of parameters for the QAOA with the unitaries associated with the edges of the 3-regular graph.  Coincidentally a 3-regular graph with 16 bits has 24 edges so this $p=4$ search was also over 160 parameters.  The best of ten gave an approximation ratio of 0.9534 which shows that the degradation in moving from the original graph to the grid was small, at least in this example.

The simulation took advantage of the fact that with a classical computer simulating a quantum computer, after building the state \eqref{eq:29}, we can simply calculate the expected value of $C$ without performing measurements.  In the algorithm given as pseudo code,  running on a quantum computer there would be $R$ repetitions of the building of the state \eqref{eq:29} with fixed angles followed by measurements producing $R$ strings $z$ and values $C(z)$.  Say $R$ is 100 to get good statistics.  We tracked the probability of finding a string $z$ which has $C(z)$ equal to the global maximum of $C$.  We found that this probability went above 1/100 long before the expected value of $C$ achieved the stopping criterion.  This will only be of potential asymptotic computation value if the standard deviation of $C$ in the state \eqref{eq:29} scales with $n$.  We know that with $p$ fixed (that is not growing with $n$) for 3-regular MaxCut the standard deviation grows only as $\sqrt{n}$.  However with $p$ of order $\sqrt{n}$ the argument in \cite{11} no longer applies.

\subsection{Warm Starts}
We are proposing using a quantum computer in a certain way to find approximate solutions to combinatorial search problems.  We have offered no evidence that we can find such solutions more efficiently than the best classical algorithms.  We now explore the possibility of using the result of the best classical algorithm as the starting point of the quantum optimization.  In this way the objective value returned by the quantum algorithm is no worse than the one returned by the best classical algorithm.  For definiteness stick with the $X$ and $ZZ$ unitaries on the grid as just discussed and a bit flip symmetric objective function $C$. Run a classical algorithm to find an approximate solution of $C$. Call this string $w$.  Now run the quantum algorithm as described in Section \ref{sec:3-1}, using $C_{\text{quad}}$,  to find a set of parameters that give the entangled state (\ref{eq:w-cat}). Then return to using the objective function $C$.  Start the optimization with the angles that produce the state $\ket w + \ket{\bar w}$. The expected value of  $C$ is $C(w)$ which is what the classical algorithm gave.  The quantum algorithm is designed to go uphill from there.  What gives us hope that this might be advantageous?  The starting parameters are the maximum of the expected value of the objective function $C_{\text{quad}}$.  But this need not be a local maximum of the objective function $C$ so we should be able to go up hill from there.  We now describe a small numerical experiment that supports this view. 

We tossed a random 3-regular graph with 16 vertices and used the MaxCut objective function. The maximum number of satisfiable edges was 20.  Since there are 24 edges a random string can be expected to satisfy 12 edges.  For our warm start we took one of the strings that satisfies 17 edges.  Call this string $w$ and its bit flip complement $\bar{w}$.  Working on the 4 by 4 grid with $p=4$ we ran a simulation of the quantum algorithm to maximize the objective function  \eqref{eq:35}.  This produced a set of 160 angles that make the state \eqref{eq:29} (very nearly) equal to the state \eqref{eq:w-cat}.  We then returned to the MaxCut objective function for the 3-regular graph but again used unitaries that depend on the 4 by 4 grid.  Starting our Nelder Mead outer loop routine requires 161 sets of 160 angles. For one of these we took the set of angles just found so that one of the starting sets of angles produces a state with an expected value of C very close to 17. The Nelder Mead optimizer found angles that produced a state with objective function value very near 20, that is, with an approximation ratio very near 1.  This is better performance than we typically observed with 161 random inputs to the Nelder Mead optimizer.   The takeaway message here is that the simulation moved beyond the value of the warm start and did not get stuck in a local minimum.

For a combinatorial search problem such as MaxCut there are classical approximation algorithms with worst case performance guarantees.  For example without any restriction on the form of the graph the Goemans Williamson algorithm guarantees an approximation ratio of 0.878.  Improving this to $0.878 + \epsilon$ for any $\epsilon > 0$ would imply that the Unique Games Conjecture is false \cite{34}.  This offers a complexity theoretic context for perhaps understanding why no improvement on GW has been found to date.  However the quantum domain is less explored than the classical and it seems worthwhile to try to run quantum computers at this type of boundary.  We can start the quantum algorithm at parameter values that give the GW ratio and hope to go uphill.   The improvement in the value of the cost value must scale with $n$ to be significant. Of course with a fixed number of qubits it will never be possible to prove an asymptotic result.  Still we can run quantum computers in this region and see what happens.

\section{Discussion and Outlook}

Sometime in the future we will have an error corrected gate model quantum computer with enough logical qubits to run quantum algorithms that can outperform classical algorithms for useful tasks.  In the near term we will have quantum computers with enough qubits to be able to produce distributions that can not be replicated by the largest existing classical computers.  As these near term quantum computers are built we need to ask how best to explore their computational power beyond the fact that they operate in a realm where classical simulation of the quantum process is not available.  Can we program these devices to  perform computationally interesting tasks?

In this paper we describe a general approach to running gate model quantum computers with fixed architecture that determines which qubits are coupled to which.  We imagine creating quantum states using only the gates native to the hardware and with the circuit depth limited by the experimental reliability of each gate.  Within this framework the quantum computer acting on a given input state produces an output state that depends on the full set of parameters that control the gates.  We then imagine performing a measurement on the state to get the value of a objective function or the expected value of a quantum Hamiltonian. Using the observed value we go back and adjust the parameters in an attempt to drive the objective function uphill or the energy down.  For certain problems it might be possible to determine an efficient strategy for choosing parameters.  Or it may be that we need to treat the quantum computer as a black box which outputs observed values of an objective function given input parameters and the optimization of the parameters is approached using a general purpose optimizer.  Without being specific about which problems are most likely to benefit from this kind of quantum neural net, we see this general approach as encompassing almost any quantum algorithm that can be run on a quantum computer without error correction or compilation. We hope to test this approach on the next generation of gate model computers as they become available.

\subsection{More or Fewer Parameters}

We began our discussion of working on a fixed qubit architecture by looking at the problem MaxCut.  In this case the objective function operator is a sum of $ZZ$ terms.  In the original version of the QAOA this operator appears in the unitaries that drive the evolution as well as single qubit $X$ rotations.  We imported these types of operators to the grid architecture.  At first we looked at the lowest depth version of the algorithm and restricted all of the $ZZ$ terms to have the same parameter and each $X$ rotation to be by the same angle. So we began with two parameters. We then advocated opening up the search space to include different angles for each $ZZ$ coupling and for each $X$ rotation. Moving to higher depth, these parameters can be different each time the corresponding operator is applied.  We can go further still and imagine any single qubit unitary (3 parameters) and any two qubit unitary (15 parameters) being applied as long as the unitaries are between qubits that are connected by the architecture and within the range of experimental capabilities. 

Opening up the search space has the advantage that it may allow us to access more favorable computational pathways. But it has the disadvantage of possibly making discovering these advantageous pathways more difficult.  For example suppose we introduce single $Y$ rotations.  Acting with the gate $\exp (\pm i Y \tfrac{\pi}{4})$ on a single $\ket{+}$ state gives the computational basis state $\ket{0}$ or $\ket{1}$  depending on the sign.  This means that if we just use a depth one algorithm with only these gates acting on the state $\ket{s}$ we can produce any computational basis state.  With these gates the solution to any NP-hard problem is in the domain of a depth one circuit with $n$ parameters.  But this procedure uses only product states and is essentially equivalent to replacing bits by classical rotating spins and presumably requires exponential time in worst case.  So introducing these gates may not facilitate finding good parameters in an opened up search space. Our hope, and it is really only a hope, is that some form of restricted entangling quantum procedure will allow us to find good approximate solutions to combinatorial search problems that evade our best classical algorithms.  Future algorithm designers will have to decide whether to use more or fewer parameters as part of their search strategies.

\subsection{Ground State Energy}
Variational quantum approaches have been suggested as a way to find the ground state energy of a quantum system. Finding states that are entangled superpositions is a natural task for a quantum computer. We imagine that the algorithm designer has mapped the system to be simulated to physical qubits of the hardware and seeks the minimum energy of the associated Hamiltonian. In the context of our work we imagine constructing quantum states via a sequence of unitaries that are experimentally available.  Once an output state has been produced the Hamiltonian must be measured. To facilitate measurement the Hamiltonian is decomposed as a sum of products of single Pauli operators. To estimate \eqref{eq:4} one can get the expected value of each single Pauli product by repeated measurements of freshly prepared output states. This introduces a factor into the computational cost which is the number of terms in the decomposed Hamiltonian. Reducing the number of qubits, gate operations and measurements necessary to simulate the original system is an active area of study. 

Note that a difference between Hamiltonian simulation and combinatorial optimization is that in the simulation case the Hamiltonian is a sum of non-commuting terms that are non-diagonal in the computational basis. In the case of combinatorial optimization, the objective function is evaluated by directly measuring in the computational basis and then classically computing $C(z)$. Note that the evaluation of $C(z)$ on a conventional computer allows for greater freedom in the set of objective functions than those that are tied to the connectivity of the hardware. As the algorithm proceeds the quantum computer may produce a string with a high value of the objective function and terminate early. It is not clear how this might carry over to the Hamiltonian case.

\subsection{Error}
There are two sources of error which need to be discussed.  First there is control error where the actual applied gates do not correspond perfectly to the desired gates.  If this is due to systematic error, the optimization strategy outlined above will not be sensitive to this kind of error.  If you are walking on terrain and want to go uphill it is not important to know your coordinates as long as you can sense which way is up.  However if the control error is different each time the quantum computer is called, this may be more problematic.  There can also be decoherence which leads to degradation of the fidelity between the actual output state and the desired state.  If the fidelity is high we can be sure that the output objective function value will be close to the ideal. But it is possible even with low fidelity that the output objective function value will be reliable (and it is easily checked). It is also possible that the outer loop optimizer will avoid computational pathways where performance is degraded due to decoherence. It may be possible to study the effects of error by using simple objective functions and setting parameters to create states with calculable objective function values and then comparing with experimental observation.

\subsection{Numerics}
Some of our numerical work was at 16 qubits working on a 4 by 4 grid at $p=4$ building quantum states that depend on 160 parameters.  The enveloping classical outer loop was called thousands of times to create a quantum state with a high value of the objective function.  Running on an actual quantum computer would require many measurements to produce good estimates of the expected value of the objective function so the thousands of simulation calls might correspond to hundreds of thousands of calls to a quantum computer.  (This is an overestimate if an early measurement produces a string with a sufficiently high value of the objective function.)  And this is all to find the maximum of a function that depends on 16 bits where brute force search over $2^{16}$ inputs is almost instantaneous on a classical computer.  So our numerics only demonstrate that these strategies can work, not that they are better than classical.  As we go to higher bit number we presume that classical methods become unwieldly for computationally difficult problems.  Still we remain optimistic that the quantum neural network will have pathways towards good solutions through routes unavailable to classical algorithms. Perhaps these can be used to solve problems more efficiently than classical algorithms.  Perhaps we will see signs of this running actual devices at scores of qubits.

\subsection{Future Strategies}
We have outlined a framework for a large class of algorithms which can be run on near term quantum computers.  We explored combinatorial optimization with a specific objective function but this approach can be applied to almost any optimization problem.  What is required are strategies for finding good parameters to drive the optimization.  Here we can imagine using warm starts with the angles initially chosen to guarantee classical performance thresholds. Or we might wish to start with restrictions on the search space to make the search more efficient and then use the result of the restricted search to seed an opened up search for example as performed in coarse-to-fine strategies. Another idea is to restrict a subset of the gates such that they correspond to known quantum subroutines such as phase estimation \cite{35} or amplitude amplification \cite{36}. The success of our proposed approach relies on having an outer loop classical optimizer tailored to generating suitable gate parameters. An attractive choice to explore is deep reinforcement learning which recently achieved impressive results \cite{37,38}. Of course ultimate success would be to find strategies  that guarantee certain performance levels in advance of running on the quantum computer. But as actual devices become available we should run them to help discover these strategies or to give us heuristics that convince us that quantum computers have power beyond classical.

\section{Acknowledgements}
E.F. and H.N. would like to thank Ryan Babbush, Aram Harrow, Sergio Boixo for stimulating discussions throughout this project. We thank Dave Bacon, John Martinis and Masoud Mohseni for comments on the manuscript and Charles Suggs for technical support. This work is supported by the National Science Foundation under grant contract number CCF-1525130.

%\appendix
\section*{Appendix}
\renewcommand{\theequation}{A.\arabic{equation}}
\setcounter{equation}{0}
%\numberwithin{equation}{section}
%\section{Appendix}
Here we show how we evaluate \eqref{eq:contribution} for the four cases pictured in figure 2. We can write \eqref{eq:contribution} as 
\be
-\smhalf \bra{s} \exp\, (i\gamma G) \exp\, (i\beta B) \, Z_i Z_j\,  \exp\, (-i\beta B) \exp\, (-i \gamma G) \ket{s} . \label{sec:app1}
\ee
The single qubit rotations give
\be
-\smhalf \bra{s} \exp\, (i\gamma G)\ [Z_i \cos 2 \beta + Y_i \sin 2\beta] [Z_j \cos 2\beta + Y_j \sin 2\beta]\ \exp (-i\gamma G) \ket{s}. \label{sec:app2}
\ee
Inserting $\exp\, (-i\gamma G) \exp\, (i\gamma G)$ in the middle we see that we need to evaluate
\be
\exp\, (i\gamma G)\, [Z_i \cos 2 \beta + Y_i \sin 2\beta] \exp\, (-i\gamma G)\label{sec:app3}
\ee
and the same thing with $i$ replaced by $j$. $G$ is a sum of products of pairs of $Z$ operators and to evaluate \eqref{sec:app3} we need only consider pairing of the form $Z_i Z_k$ where $\vev{ik}$ is an edge in the grid. Note that $k$ might be equal to $j$. Now
\begin{alignat}{2}
\exp \left(-i\frac{\gamma}{2}\ Z_i Z_k\right)  Z_i \exp \left(i\frac{\gamma}{2}\ Z_i Z_k\right)& = Z_i \nonumber\\[1.5ex]
\exp \left(-i\frac{\gamma}{2}\ Z_i Z_k\right) Y_i \exp \left(i\frac{\gamma}{2}\ Z_i Z_k\right) & = Y_i \cos \gamma -X_i Z_k \sin \gamma \label{sec:app4}\\[1.5ex]
\exp \left(-i\frac{\gamma}{2}\ Z_i Z_k\right)  X_i \exp \left(i\frac{\gamma}{2}\ Z_i Z_k\right) &= X_i \cos \gamma + Y_i Z_k \sin \gamma .\nonumber
\end{alignat}
By repeated applications of \eqref{sec:app4} we can expand \eqref{sec:app3} out as well as \eqref{sec:app3} with $i$ replaced by $j$.
We take the expectation of the product in the state $\ket{s}$. Since $\ket{s}$ is an eigenstate of each $X_\ell$ with eigenvalue 1 and $\bra{s} Y_\ell \ket{s} = \bra{s} Z_\ell \ket{s} = 0$ for each $\ell$, the only terms that survive are those which are products of $X_\ell$'s with no unpaired $Y_\ell$ or $Z_\ell$.

Consider case $0$ which is depicted in Figure 2.0. Label the four vertices connected to $i$ as 1, 2,3,4. In \eqref{sec:app3} the only relevant piece of $G$ is $-\tfrac{1}{2} [Z_i Z_1 + Z_i Z_2 + Z_i Z_3 + Z_i Z_4]$ which is a sum of commuting terms. Consider the action of $\exp \left(-i\frac{\gamma}{2} Z_i Z_1\right)$ using \eqref{sec:app4}. Since $\bra{s}Z_1\ket{s}=0$ we need only keep terms proportional to $Z_i$ and $Y_i$. Repeating with 2, 3, and 4 we still only end up with terms proportional to $Z_i$ and $Y_i$, but the expected value of these terms in the state $\ket{s}$ is 0. So we see that for case 0, the expression \eqref{sec:app1} is 0. 

We now turn to case 1. Vertex $i$ has four neighbors, one of which is $j$ and the other three are not neighbors of $j$.  Apply \eqref{sec:app4} three times to \eqref{sec:app3} with the three neighbors of $i$ other than $j$. This gives
\be
Z_i \cos 2 \beta + Y_i \sin 2 \beta \cos^3 \gamma .\label{sec:app5}
\ee
since the expectation of $Z_\ell$  in $\ket{s}$ is 0 for these three neighbors. Now apply $Z_i Z_j$ using \eqref{sec:app4} to get
\be
Z_i \cos 2\beta + \left(Y_i \cos \gamma - X_i Z_j \sin \gamma \right)\ \sin 2 \beta \cos^3 \gamma .\label{sec:app6}
\ee
Repeating this with $i$ replaced by $j$ gives 
\be
Z_j \cos 2 \beta + \left(Y_j \cos \gamma - X_j Z_i \sin \gamma\right) \ \sin 2 \beta \cos^3 \gamma . \label{sec:app7}
\ee
Taking the expectation in $\ket{s}$ of the product of \eqref{sec:app6} and \eqref{sec:app7} gives
\be
-2 \cos 2 \beta \sin 2 \beta \ \sin \gamma \cos^3 \gamma\label{sec:app8}
\ee
which when plugged into \eqref{sec:app1} gives
\be
 \smhalf \sin 4\beta \sin \gamma \cos^3\gamma
\ee
which is \eqref{fsmall}.

In Cases 2 and 3 there is no term $Z_i Z_j$ in $G$ but vertices $i$ and $j$ do have neighbors in common (two in case 2 and one in case 3) and the non-zero contributions to the expectation come from pairing the $Z_k$ from these common neighbors in the expressions of \eqref{sec:app4}, which leads to \eqref{eq:17} and \eqref{eq:18}.

%\bibliographystyle{plain} 
%\bibliographystyle{unsrtnat}
%\bibliography{Fixed_Qubit_Layout}
%\bibliography{citations/fixedqubit}

%\nocite{*} % adds all entries in the bib file to the bibliography
%\printbibliography

\end{document}